# Electric Current Induced Ultrafast Demagnetization

Richard B. Wilson[a,b,1,], Yang Yang[a,2], Jon Gorchon[3], Charles-Henri Lambert[4], Sayeef Salahuddin[3,4], Jeffrey Bokor[b,3,4].

[1]Department of Mechanical Engineering and Materials Science and Engineering Program, University of California, Riverside, CA 92521, USA

[2]Department of Materials Science and Engineering, University of California, Berkeley, CA 94720, USA

[3]Lawrence Berkeley National Laboratory, 1 Cyclotron Road, Berkeley, CA 94720, USA

[4]Department of Electrical Engineering and Computer Sciences, University of California, Berkeley, CA 94720, USA

[a]denotes equal contributions

[b]To whom correspondence should be addressed.   Email: rwilson@engr.ucr.edu, jbokor@berkeley.edu.

**Abstract**

We report the magnetic response of Co/Pt multilayers to picosecond electrical heating. Using photoconductive Auston switches, we generate electrical pulses with 5.5 picosecond duration and hundreds of pico-Joules to pass through Co/Pt multilayers. The electrical pulse heats the electrons in the Co/Pt multilayers and causes an ultrafast reduction in the magnetic moment. A comparison between optical and electrically induced demagnetization of the Co/Pt multilayers reveals significantly different dynamics for optical vs. electrical heating. We attribute the disparate dynamics to the dependence of the electron-phonon interaction on the average energy and total number of initially excited electrons.

## I. Introduction

The pioneering observation of ultrafast demagnetization in ferromagnetic nickel following optical irradiation[1] has led to the discovery of a broad range of extraordinary magnetic phenomena. Laser irradiation of magnetic metals can launch precessional modes at frequencies ranging from a few to hundreds of GHz [2,3], drive ultrafast magnetic phase transitions [4], and generate enormous pure spin-currents [5-11]. Optical irradiation of ferrimagnetic systems such as GdFeCo and TbFeCo can result in an ultrafast reversal of the direction of magnetization[12-14]. Several recent studies have observed the response of magnetic metals to free-space THz radiation[15,16].

Despite this broad array of discoveries, the microscopic mechanisms that enable the sub-picosecond quenching of the magnetization in magnetic metals following ultrafast heating are unclear[17,18]. One aspect of optically induced ultrafast demagnetization that remains under debate is whether the initially *nonthermal* distribution of electrons is an important driver of ultrafast magnetic phenomena[17,19-21]. In the first hundred femtoseconds following laser irradiation, electrons are nonthermal, i.e. Fermi-Dirac statistics provides a poor description of the excitation energies [22]. Several studies have predicted the initially nonthermal distribution impacts ultrafast demagnetization because electronic scattering rates depend on both the average energy and total number of electronic excitations[19,23]. The average energy and total number of excitations can also impact transport phenomena, which may be important in the ultrafast demagnetization in metal multilayers [24]. However, the lifetimes of eV-scale electronic excitations are often only tens of femtoseconds[18]. Demagnetization typically occurs over hundreds of femtoseconds[25]. Therefore, most models assume that highly excited electronic states can be disregarded when modelling magnetization dynamics[20] and treat the electron distribution as thermal on all time-scales.

Our work experimentally demonstrates that the initially nonthermal distribution of electrons can strongly impact optically induced ultrafast magnetization dynamics by modifying the rate of energy transfer between electrons and phonons. We deposit roughly equal amounts of energy into the electrons of a magnetic film with either a 2.6 picosecond optical pulse or 4 picosecond electrical pulse. Optical heating deposits energy by exciting a few electrons ~ 1.5 eV above the Fermi-level. In contrast, electrical heating simultaneously excites many electrons to only a few meV above the Fermi level. These differences in the initial electron distribution cause significant differences in the magnetization dynamics. The nonthermal electron distribution that optical irradiation excites transfers energy to the phonons at a significantly reduced rate in comparison to the distribution of electrons excited by electrical heating.

## II. Methods

We excite electrical pulses with a 5.5 ps duration on a coplanar waveguide structure (CPW) using photoconductive Auston switches (Fig. 1). Additional details concerning device properties and fabrication are in Ref. [26]. To bias the photoconductive switch during operation, we connect one side of the CPW device to a DC voltage source. Upon optical irradiation of the biased photoconductive switch with an 810 nm laser, a transient electrical pulse with a FWHM of 5.5 ps is generated and propagates along the CPW (Fig. 1d). The current profile $I(t)$ is measured with a Protemics THz detector[26,27]. The energy carried by the electrical pulses, $\int I^2(t) Z_0 dt$, ranges from 1 to 200 pJ for DC biases across the photoswitch between 10 and 80 V. The impedance $Z_0$ of the waveguide is ~60 ohms. A small section of the CPW center line is made of a ferromagnetic thin film, see Fig. 1c. Upon passing through the ferromagnetic wire, the electrical pulse deposits part of its energy via Joule heating, thereby inducing ultrafast demagnetization. Since the power scales with $I^2$, the 5.5 picosecond current pulse corresponds to a ~4 ps heat pulse.

Optical ultrafast demagnetization experiments are typically performed with laser fluences between 0-10 J m$^{-2}$ [10,28,29], corresponding to irradiation of the film with 0.1 to 1 nJ of energy across a 100 µm² region. Our CPW device delivers similar energy densities with an electrical pulse to a ferromagnetic wire. At a distance of 0.5 mm from the photoconductive switch, the center line width of the CPW and gap distance between the center line and ground are tapered down from ~30 um to ~5 um over 0.6 mm. The ratio between the centerline width and gap distance is constant in order to keep the waveguide impedance constant at 60 ohms. In the narrowed region of the CPW, a 5 um long section of the center line is made out of a thin film of a ferromagnetic metal. We are able to deliver electrical pulses with energies as high as ~200 pJ to a 25 µm$^2$ ferromagnetic thin film, i.e we can deliver up to ~8 J m$^{-2}$ of electrical fluence. Only a fraction of the incident energy is absorbed via Joule heating, e.g. ~10 to 30%. The amount of electrical energy absorbed depends on the resistivity and dimensions of the ferromagnetic wire. We use a multilayer calculation to compute the absorption[26].

In addition to the photocurrent across the device, a constant but small dark current flows across the device in the absence of laser illumination of the photoswitch, see Fig. 1d. In the experiments, we describe below, the dark current is less than 20 µA because the bias voltage is kept below 60 V. The heat-current on the magnetic device due to the dark current is less than 2 mJ m$^{-2}$ and has no impact on the experiments.

We characterize the magnetization response of the CoPt samples to heating via time-resolved measurements of the polar magneto-optical Kerr effect (TR-MOKE). We modulate the pump beam with an electro-optic modulator at 1 MHz and use lock-in detection to monitor small changes to the magneto-optic response of the sample. The duration of the probe laser pulse is 0.3 ps, much shorter than the 2.6 ps pump pulse. The optical pump and probe beams possess different pulse durations because of dispersion from the electro-optic-modulator that the pump beam passes through. Optical pulse durations are determined with an APE autocorrelator. The 1/e radius of the pump beam focused on the sample is ~20 µm. The 1/e radius of the probe beam is ~ 1.5 µm. The spot-size is determined in two ways. First, we use the knife-edge method. Second we use a CCD camera image of the beam profile. Both agree to within 10%.

The experimental setup includes an integrated microscope that uses bright field imaging to monitor the pump and probe beams on the sample surface. The vibration isolation provided by our optics table ensures sub-micron stability so that the spatial jitter experienced by the laser beams is much less than the micron scale features of the devices.

In addition to performing electrical demagnetization experiments, we perform optical demagnetization experiments by altering two things in our experimental setup. Instead of focusing the pump beam on the photoconductive Auston switch, we overlap the pump and probe beams on the sample. Additionally, the optical demagnetization experiments are not performed on the 5 µm x 5 µm section of the ferromagnetic wire that we pass the electrical current through because the pump beam radius of 20 µm is much larger than 5 µm.  Instead, we move the pump and probe beam to a separate area of the sample where a large section of the Co/Pt multilayer film remains un-patterned. We confirmed that magnetization dynamics we observe are not sensitive to the spatial location of the film by performing optical pump/probe measurements at four different spatial locations. These different spatial locations included both patterned regions and un-patterned regions of the sample.  In all cases, the magnetization dynamics on ten picosecond time-scales are identical. To prevent optical artifacts in our signal, we implement a "two-tint" approach in our experiments by red/blue shifting the pump/probe beams with sharp-edged optical filters. Without optical filters, the laser pulse has a 50 nm bandwidth centered at 810 nm. Upon insertion of the long pass filter on the pump path, the bandwidth is 20 nm and is centered at 823 nm.  Upon insertion of the short pass filter on the probe beam, the bandwidth is 12 nm centered at 795 nm.

### III. Results: Electrical versus Optical Demagnetization

We performed both optical and electrical ultrafast demagnetization experiments on two Co/Pt multilayers (Fig. 2).  The geometry of the first and second film are (3 nm Ta / 15 nm Pt / [0.7 nm Pt / 0.6 nm Co] x 8 / 5 nm Pt), and (1 nm Ta / 1 nm Pt / [0.7 nm Pt / 0.6 nm Co] x 8 / 1.7 nm Pt), respectively. Below, we refer to these as the Pt/CoPt and the CoPt sample, respectively.

Figures 2 and 3 show the response of the two samples to ultrafast heating of the electrons via optical (Fig. 2) and electrical pulses (Fig. 3a and 3b). The absorbed optical fluence for the Pt/CoPt and CoPt samples is 0.2 J m$^{-2}$ and 0.7 J m$^{-2}$, respectively. We use a 50 V bias voltage in the electrical demagnetization experiments shown in Figs. 2 and 3. All demagnetization curves are normalized by the demagnetization at 10 ps to facilitate comparisons.

At large heating fluences, nonlinear effects are known to impact the magnetization dynamics.  An example of such a nonlinear effect is the increase in the demagnetization time-scale that occurs if the fluence is sufficient for the per-pulse temperature rise to approach the Curie temperature[20,25,28]. Here, we intentionally use small fluences in order to exclude nonlinear effects from the dynamics and simplify analysis. The peak demagnetization in the optical and electrical experiments shown in Figs. 2 and 3 is below 2%, and the peak per pulse temperature rise is less than 20 K. The peak demagnetization is calculated by normalizing the maximum transient Kerr rotation in our pump/probe experiments by the static Kerr rotation. The maximum temperature rise is calculated with a thermal model[28], which is described in detail in the next section. We experimentally verify that our experiments are in a linear regime by performing optical and electrical experiments across a range of absorbed fluences between 0 and 0.7 J m$^{-2}$. The shape of the dynamics does not depend on the bias voltage in the electrical experiments, or the laser energy in the optical experiments[26].

For both electrical and optical heating, the magnetization of the Co/Pt multilayer decreases rapidly, se Figs. 2 and 3. However, clear differences exist for the two types of heating. We attribute the differences to differences in the initial distribution of excited electrons. In Supplemental Material[26], we rule out significant contributions to the electrical demagnetization signal from effects such as differences in pulse duration[30], Oersted fields that accompany the transient electrical pulse[31], the spin Hall effect from strong spin orbit coupling in the Pt[32], or optical state blocking effects[33].

Optical irradiation excites electrons between 0 and 1.55 eV above the Fermi level and the initial distribution is nonthermal, i.e. can't be described with Fermi-Dirac statistics [34,35]. In contrast to optical heating, when electrons are electrically heated their energies only increase a few meV. The largest longitudinal electric field that occurs in the ferromagnetic wire during our experiments is $j_{max}/\sigma \approx 4 \text{ MV m}^{-1}$, where $j_{max}$ is the maximum current density and $\sigma$ is the electrical conductivity of the ferromagnet. Assuming a scattering time of ~ 30 fs, a value typical for transition metals [36], the average increase in kinetic energy of an electron due to acceleration in the electric field prior to scattering is $\Delta E \approx (eE\tau)^2/(2m_e) \approx 1 \text{ meV}$, where $m_e$ is the mass of an electron. Because $\Delta E \ll k_B T$, the distribution of excited electrons is thermal. Therefore, by comparing the response of CoPt and Pt/CoPt to optical vs. electrical heating, we are directly probing the impact of an initially nonthermal vs. thermal electron distribution on the magnetization dynamics.

Demagnetization of the CoPt following optical heating, as shown in Fig. 2, displays "type I" dynamics [20]. The sample demagnetizes during laser irradiation, followed by a smaller increase in the magnetization as the electrons and phonons thermalize. Our "type I" categorization agrees with prior studies of Co/Pt [37], whose large spin-orbit coupling is credited with abnormally strong coupling between electronic and spin degrees-of-freedom. Magnetization dynamics without a recovery in the magnetization in the picoseconds following irradiation are "type II" dynamics [20]. (The category of "type II" dynamics also includes observations of demagnetization on multiple time scales, as is observed for Gd[20].) In contrast to the "type I" dynamics displayed following optical irradiation, the magnetization of neither CoPt or Pt/CoPt display a significant recovery in the picoseconds following heating (Fig. 3).

IV.   Thermal Model Analysis

After energy is added to the electronic system via optical or electrical pulses, the electrons transfer energy to other degrees of freedom via electron-phonon scattering, and scattering between the electrons and spin degrees-of-freedom [8]. Scattering between the electronic and spin degrees-of-freedom of the metal increases populations of spin excitations, e.g. magnons, spin-density fluctuations, and Stoner excitations [28,38]. As a result, ultrafast heating of the electrons rapidly reduces the total magnetization (Fig. 2).

We model the redistribution of energy from optically excited electrons to phonons and spin degrees-of-freedom with a phenomological three temperature model [1,28]. The three temperature model accounts for the ability of electrons, spins, and phonons to store different amounts of energy per degree of freedom through distinct electron, spin, and phonon temperatures: $T_e$, $T_s$, and $T_p$. We compute the absorption of energy in the metal from the laser pulse with a multilayer optical calculation that predicts the absorption profile vs. depth. The model accounts for thermal diffusion across the Pt/CoPt and CoPt multilayers by electrons by including diffusion terms in the heat equation. The electronic thermal conductivity is fixed via electrical resistivity measurements and the Wiedemann-Franz law. Four-point measurements yielded electrical resistivities for the CoPt/Pt and CoPt samples of 4 and 5 x $10^{-7}$ Ωm, respectively. The electronic heat-capacity is set based on first-principles calculations of the electronic density of states of Pt. The volumetric phonon heat-capacity is fixed using experimental values from literature. Further details of the thermal model and multilayer absorption calculations are provided in Ref. [26].

We emphasize that while the thermal model has many parameters, its predictions are only sensitive to the electron-phonon energy transfer coefficient, the phonon heat-capacity, and heat pulse duration because of the picosecond duration of the pulses. The picosecond time-scale for heating in both our optical and electrical demagnetization experiments is much greater than the electron-spin relaxation time of $\tau_{es}$ ~ 40 fs in Co/Pt multilayers, and the electron-phonon relaxation time, $\tau_{ep}$, which is typically on the order of a few hundred femtoseconds in transition metals. This does not imply the electrons, phonons, and spins are well described by a single temperature, i.e. that the various excitations are all in thermal equilibrium. The electron-spin, and electron-phonon time-scales describe how quickly the electron and spin temperatures can reach a quasi-steady-state condition where the inflows and outflows of heat are roughly equal. For the spins, this implies $T_e \approx T_s$, because the spins are not directly heated by the laser, and are not strongly coupled to the phonons [39]. Therefore, for picosecond heating, our three temperature model becomes a two temperature model. Alternatively, the picosecond heating imposes a different condition on the electron and phonon temperatures. Because the picosecond heating is much greater than the electron-phonon relaxation time, the electron temperature will be such that $q \approx g_{ep}(T_e - T_p)$, where $q$ is the time dependent electronic or optical heating of the electrons. Therefore, the predictions of the thermal model for both optical or electrical demagnetization are only sensitive to three parameters. The heating profile vs. time $q(t)$, the electron-phonon energy transfer coefficient $g_{ep}$, and the phonon heat-capacity $C_p$. The sensitivity to the phonon heat-capacity arises due to the fact that the phonon-temperature is evolving in time.

The heating profile $q(t)$ and the value of $g_{ep}/C_p$ determines the shape of the demagnetization curve. The heating profile for the optical experiments is accurately measured using an APE autocorrelator that has fs resolution. The heating profile for the electrical experiments is accurately measured by measuring the electric field vs. time of the current pulse with a Protemics THz detector. The Protemics detector possess sub-ps resolution[27]. The phonon heat-capacity is set to 2.85 and 3.1 MJ m$^{-3}$ K$^{-1}$ for the Pt and [Co/Pt] multilayers based on literature values, respectively.

We fix $g_{ep}$ in our thermal model using the scattering theory original derived by Allen[40]. According to scattering theory[40], $g_{ep} \approx \pi \hbar k_B D_f \lambda \langle \omega^2 \rangle$, where $\hbar$ is the reduced Planck's constant, $k_B$ is Boltzmann's constant, $D_f$ is the density of states at the Fermi level, $\lambda$ is the electron-phonon coupling constant in the Eliashberg generalization of BCS theory, and $\langle \omega^2 \rangle$ is the second frequency moment of the phonons. We approximate $\langle \omega^2 \rangle$ by assuming a Debye density of states, $\langle \omega^2 \rangle \approx 0.6 \cdot k_B^2 \Theta_D^2 / \hbar^2$, where $\Theta_D$ is the Debye temperature. For Pt, $D_f \sim$ 9 x 10$^{47}$ J$^{-1}$ m$^{-3}$ [41], $\lambda = 0.66$ [42], and $\Theta_D \approx 240$ K. Therefore, theory predicts $g_{ep} \approx$ 1.5 x 10$^{18}$ W m$^{-3}$ K$^{-1}$. We use the theory prediction for $g_{ep}$ of Pt only because no experimental measurement of $\lambda$ exists for Co. First-principles calculations suggest $g_{ep}$ is higher for Co than Pt,[43,44] therefore 1.5 x 10$^{18}$ W m$^{-3}$ K$^{-1}$ can be viewed as a theoretical estimate of the lower limit for $g_{ep}$ in our layers.

To simplify comparisons between the model predictions to the experimental data, we normalize the predicted demagnetization at all time delays by the measured demagnetization at 10 ps. This final normalization step removes the sensitivity of the model's predictions to parameters that determine the magnitude of the demagnetization curve, i.e. the volumetric phonon heat capacity of the metals, the energy absorption coefficient, and the temperature dependence of the magnetization.

Three temperature model predictions are in excellent agreement with the electrically induced demagnetization data, see Fig 3. However, the three-temperature model predictions are in poor agreement with the optical experiments, see Fig. 2. In order to achieve good agreement between the thermal model and the optical demagnetization data for both the CoPt and Pt/CoPt sample, we must reduce the value of the electron-phonon coupling constant by half to 7 x 10$^{17}$ W m$^{-3}$ K$^{-1}$. The factor of two difference in the peak electrical vs. electrical demagnetization cannot be explained by the differences in optical vs. electrical pulse duration. The thermal model predicts that a 35% change in pulse duration from 4 to 2.6 ps will only alter the peak demagnetization by ~10%. Instead, we posit that the disagreement between the optical demagnetization data and the three-temperature model is because the three-temperature model does not account for the initially nonthermal distribution of excited electrons in the optical experiments.[35] We discuss this further in section V.

In the above analysis, we restricted our comparison between the demagnetization data and thermal model predictions to the shape of the demagnetization. Now, we compare the magnitude of the demagnetization at 10 ps delay time to the predictions of our thermal model. In Fig. 4, we plot the demagnetization as a function of the peak current of the pulse. Uncertainty in our electrical absorption calculations is ~30% due to uncertainties in the film resistivity and dimensions. In order to make predictions with the thermal model for the demagnetization, we must have knowledge of the temperature dependence for the magnetization. We set $1/M \left( dM/dT \right) \approx 10^{-3}$ K$^{-1}$ by comparing the optical demagnetization at 10 ps to the per pulse temperature rise, $hC_{tot}/F$, where $h$ is the metal film thickness, $C_{tot}$ is the total volumetric heat capacity, and $F$ is the absorbed fluence. The agreement between data and model predictions supports our conclusion that the observed ultrafast magnetic response of both samples is due to electrical heating.

## V. Nonthermal Model Analysis

Photoemission experiments suggest the nonthermal electron distribution initially excited by an optical pulse persists for tens to hundreds of femtoseconds in transition metals such as Al [35], Au [34], Ni [22], and Fe [45]. While an electron-electron equilibration time of 10 fs < $\tau_{ee}$ < 100 fs is much shorter than the picosecond time-scales of our heat pulses, that does not imply the initial nonthermal distribution has no effect in our experiments. As we demonstrate below, the relevant comparison is not between the time-scale for heating and $\tau_{ee}$. Instead, the relevant comparison is between $\tau_{ee}$ and $\tau_{es}$, or between $\tau_{ee}$ and $\tau_{ep}$. Unless $\tau_{ee}/\tau_{es} \ll 1$ or $\tau_{ee}/\tau_{ep} \ll 1$, a significant fraction of the energy transfer to spin and vibrational degrees-of-freedom from photoexcited electrons occurs while the electrons are nonthermal. Thus, unless $\tau_{ee}/\tau_{ep} \ll 1$, the effective thermal resistance between a material's electrons and phonons will depend strongly on whether the initially excited distribution is thermal or nonthermal.

To demonstrate that nonthermal heating has a significant impact on the interaction of electrons and the lattice, we consider the energy dynamics of the electrons for three situations: (1) a nonthermal distribution of Pt electrons that transfers energy to the lattice in the absence of electron-electron scattering ($\tau_{ee}/\tau_{ep} \gg 1$), (2) a nonthermal distribution of Pt electrons that transfers energy to the lattice while undergoing electron-electron scattering ($\tau_{ee}/\tau_{ep} \sim 1$), and (3) a thermal distribution of Pt electrons ($\tau_{ee}/\tau_{ep} \ll 1$). Our analysis is based on the nonthermal model described by Tas and Maris[35]. For simplicity, we assume the laser excites a nonthermal distribution of excitations that is independent of excitation energy, see Fig. 5a. We also neglect the temperature rise of the lattice. Finally, we neglect the energy dependence of the electron-phonon scattering rate. Then, Allen's theory for electron-phonon scattering predicts that excited electrons and holes transfer energy to the lattice at a rate of[35,40]

$$\dot{q} = \frac{\pi \hbar \lambda \langle \omega^2 \rangle}{2 \ln 2} , \quad (1)$$

which for Pt is 0.9 eV ps$^{-1}$. The total rate of energy transfer from all excited electrons to the lattice is

$$\dot{Q}_{ep}(t) = \int_0^\infty \dot{q} n(E,t) dE , \quad (2)$$

where $n(E,t)$ is the number of excitations due to heating. The number of excitations evolves in time due to electron-electron and electron-phonon scattering

$$\frac{dn(E,t)}{dt} = \frac{dn(E,t)}{dE} \frac{\hbar \langle \omega \rangle}{\tau_{ep}} - \frac{n(E,t)}{\tau_{ee}(E)} + 6\int_E^\infty \frac{(E-E')}{E^2} \frac{n(E',t)}{\tau_{ee}(E')} dE' + \frac{A(t)}{E_p} , \quad (3)$$

where $A(t)$ is the number of photons absorbed per second, and $E_p$ is the energy of the absorbed photons. The electron-electron scattering time for an excitation of energy $E$ above or below the Fermi energy $E_f$ is[35]

$$\tau_{ee} = \tau_0 \left(\frac{E_F}{E}\right)^2, \qquad (4)$$

where $\tau_0 = 128/\left(\pi^2 \sqrt{3}\omega_p\right)$, and $\omega_p$ is the plasma frequency. The electron-phonon scattering time is

$$\tau_{ep} = \frac{\dot{q}}{\hbar\langle\omega\rangle}. \qquad (5).$$

Using Eqs. (1-5), we calculate $E_{tot}(t) = \int n(E,t) E dE$ for Pt that results from the impulsive absorption of an energy density of 10 MJ m$^{-3}$, see Fig. 5b. This energy density is comparable to what we use in our experiments. For Pt, $E_f = 8.6$ eV, $\hbar\omega_p = 5.15$ eV, which implies $\tau_0 = 1$ fs and $\tau_{ee} = 30$ fs at $E = 1.55$ eV. The nonthermal model predicts the rate of energy-transfer from photoexcited electrons to the lattice occurs on a time-scale of 0.3 ps. For comparison, we also compute $E_{tot}(t)$ in the limit of negligible electron-electron scattering, and we include $E_{tot}(t)$ predicted by a two-temperature model with $g_{ep}$ = 1.5 x 10$^{18}$ W m$^{-3}$ K$^{-1}$, which is equivalent to the strong electron-electron scattering limit where $\tau_0 = 0$. In the absence of electron-electron scattering, a nonthermal electron distribution excited by 1.55 eV phonons transfers energy to the lattice on a time-scale of 0.6 ps. Alternatively, the two-temperature model predicts a time-scale of $C_{el} / g_{ep}$ ~ 0.15 ps. The factor of two difference in thermalization time-scale for the non-thermal vs. thermal model predictions is consistent with our analysis in the prior section.

Now, we extend our analysis to the electron energy dynamics in response to picosecond heat pulses. An implication of the small fluences we use in our experimental study is the heat induced dynamics are linear, i.e. superposition applies. Therefore, the picosecond heating in our experiment will induce dynamics that can be represented as a linear combination of the dynamics caused by a sequence of impulsive heat pulses. In the context of ultrafast magnetism, this implies magnetization dynamics from long heat pulses can be directly derived from the dynamics resulting from shorter heat pulses[30]. Superposition ensures that if the rate of energy exchange between electrons, spins, and phonons is sensitive to the initially excited distribution of electrons, this sensitivity remains regardless of the duration of the heat pulse. In Fig. 6, we show results for an energy absorption of 10 MJ m$^{-3}$ over a duration of 2.6 ps. We show calculations for the number of excitations and the total energy of all excitations $E_{tot}(t)$. The distribution of electrons remains nonthermal over the entire laser pulse duration because nonthermal electrons are continually excited to nonthermal energies. The key difference between impulsive heating (Fig. 5b) and picosecond heating (Fig. 5c and 5d) is that for picosecond heating the system is in a quasi-steady state, i.e. $dn/dt \approx 0$ in Eq. 3. Even in a steady-state condition, if the condition $\tau_{ee} \ll \tau_{ep}$ is not met at energies near the Fermi-level, the solution of Eq. 3 for $n(E)$ is sensitive to the functional form of $A(E,t)$. In short, our non-thermal model corroborates our hypothesis that thermal vs. nonthermal heating strongly impacts the energy evolution of the excited electrons.

## VI.    Discussion

In the prior two sections, we explain our experimental results with the hypothesis that exciting a nonthermal distribution of electrons influences the ability of electrons and phonons to exchange heat. In addition to influencing the rate of energy transfer to phonons, there are several ways for a nonthermal distribution to influence the demagnetization dynamics. For example, Elliot-Yafet scattering is thought to play a central role in ultrafast demagnetization[20] and depends on the total electron-phonon scattering rate. The scattering rate between electrons and spin-excitations [8], e.g. magnons, may also depend on the number of excited electrons. The high average energy of excitations in a nonthermal distribution may allow the generation of nonthermal spin excitations, e.g. Stoner excitations with sub-eV energies [22].  Finally, the rate that electrons thermalize with the lattice will indirectly impact the magnetization dynamics. The rate of energy transfer to the spin degrees-of-freedom depends on how long the electrons remain hot [28].  A faster exchange of energy between electrons and phonons favors slower demagnetization [28].

A nonthermal distribution of excited electrons can also impact how energy and angular momentum are transported, e.g. allow for superdiffusive spin and heat currents [5,6]. Therefore, in addition to an altered electron-phonon interaction, it is also possible that the significant differences we observe between electrical and optical demagnetization are partially due to superdiffusive spin transport. We note that superdiffusion and changes to e-p scattering rates are related phenomena, as electron-phonon scattering rates are an important component of super diffusive transport theory[46]. One motivator for the geometry of our two samples is to investigate the impact of superdiffusion on our results. The Co/Pt multilayer in the CoPt/Pt sample is sandwiched between Pt layers that are 5 and 15 nm thick, comparable to the spin diffusion length in Pt of ~ 8 nm. In contrast, the Co/Pt multilayer in the CoPt sample is sandwiched between only 1 and 1.7 nm of Pt. Therefore, the CoPt/Pt sample possesses a significant Pt reservoir for superdiffusive spins to be transported into, while the CoPt sample does not. The differences between optical and electrical demagnetization are similar for both samples. Therefore, we cannot conclude from the current experiments that superdiffusion is an important contributor for the differences in optical and electrical demagnetization. One possible explanation for the similar data for both samples is that the hot electron velocity relaxation length in Pt is much shorter than 8 nm. Recent observations of THz emission are consistent with a hot electron length of only ~1 nm[47].

Several prior studies have demonstrated that indirect optical excitation can induce ultrafast demagnetization. For example, Eschenlohr *et al.* reported differences in dynamics that occur following optical excitation of a Au/Ni vs. Ni sample to examine how nonthermal electron transport impacts magnetization dynamics. Similar experiments have been performed by Vodungbo *et al.* and Bergeard *et al* with Al. In these types of experiments, the average energy, spin-polarization, and number of excited electrons is altered depending on whether the energy is directly absorbed by the ferromagnet, or indirectly delivered via hot electrons from an adjacent metal layer, e.g. Au[19,48], Al[49], or Cu[50]. However, significant controversies remain concerning the interpretation of these experiments because they require modeling of how the initially excited distribution of electrons evolve in time and space after optical irradiation. Eschenlohr *et al.* use scattering rates from first principles calculations to estimate the distribution and transport of excitations into the ferromagnetic layer from an adjacent film, however their interpretation remains controversial[51,52]. Alternatively, Bergeard *et al.* suggest hot electron transport between metal layers is ballistic[50]. Other studies report results that are consistent with thermal diffusion[7,10,52]. Our electrical vs. optical heating approach is a more direct method for testing the impact that the initially excited electron distribution has on the magnetization dynamics because no sophisticated predictions for how an excited electron distribution evolves in time and space are necessary.

In conclusion, we observe rapid demagnetization in Co/Pt wires due to picosecond electrical heating. We observe large differences in the demagnetization rates of Co/Pt for optical vs. electrical heating. We attribute the large differences to the initially nonthermal vs. thermal distributions of excited electrons. The rate of scattering processes responsible for transferring energy from the electrons to the lattice degrees of freedom is strongly affected by the number and average energy of excited electrons. Prior studies have examined how nonthermal distributions impact electron-phonon interactions by comparing the value of $g_{ep}$ derived from pump/probe measurements to theory [35,53]. The values of $g_{ep}$ derived by fitting pump/probe measurements with a thermal model are often lower than theoretical predictions, presumably in order to compensate for the lower electron/phonon scattering rate while the electron distribution is nonthermal[35]. Our study provides the first direct test of how differences in the excited electron distribution impacts energy transfer. Finally, our experimental results will require a reexamination of the belief that the physics of optically induced demagnetization is well described by assuming the electron system is thermalized [20].

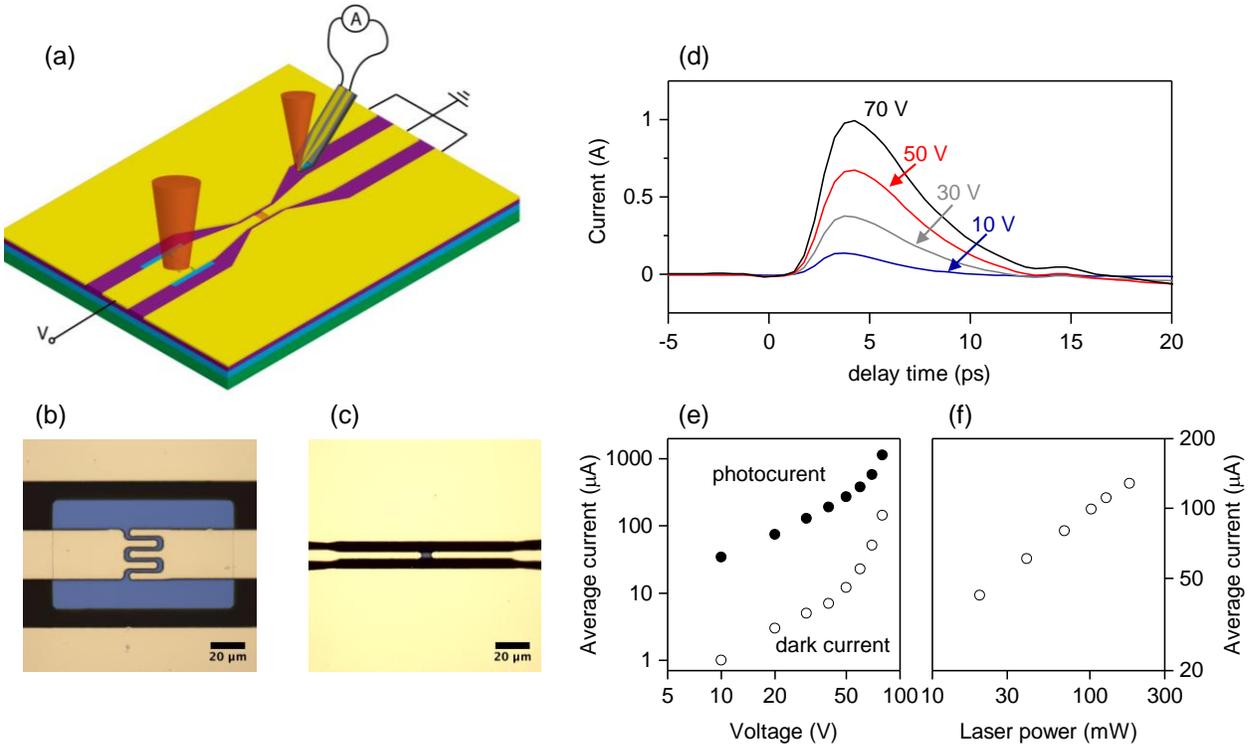

**Figure 1.** Device for electrically induced ultrafast demagnetization experiments. **a,** Schematic of the electrical demagnetization experiments. The Auston switch is illuminated with a 1.5 nJ laser pulse while biased between 10 and 80 V. The magnetization of the magnetic wire is monitored via TR-MOKE. **b,** Optical image of the Auston switch. During illumination, photoexcited carriers in the low-temperature GaAs substrate conduct current across the gap, generating a transient electrical pulse that propagates along the waveguide towards a ferromagnetic section of the centerline. **c**, Optical image of the CoPt section of the waveguide. **d,** Temporal profile of the current pulse generated by the photoswitch, as measured with a Protemics probe positioned between the photoswitch and the CoPt wire. **e,** Average current across the device with 160 mW of laser power irradiating the photoswitch. The filled circles correspond to measurements while the photoswitch was irradiated with 1.5 nJ laser pulses at a rate of 80 MHz (photocurrent). Open circles are current measurements on the device without laser irradiation (dark current). A rapid increase in darkcurrent occurs as the bias voltage across the photoswitch approaches the breakdown voltage of the device, ~90V. **f**, Dependence of the average current on average laser power with a bias voltage of 30 V.

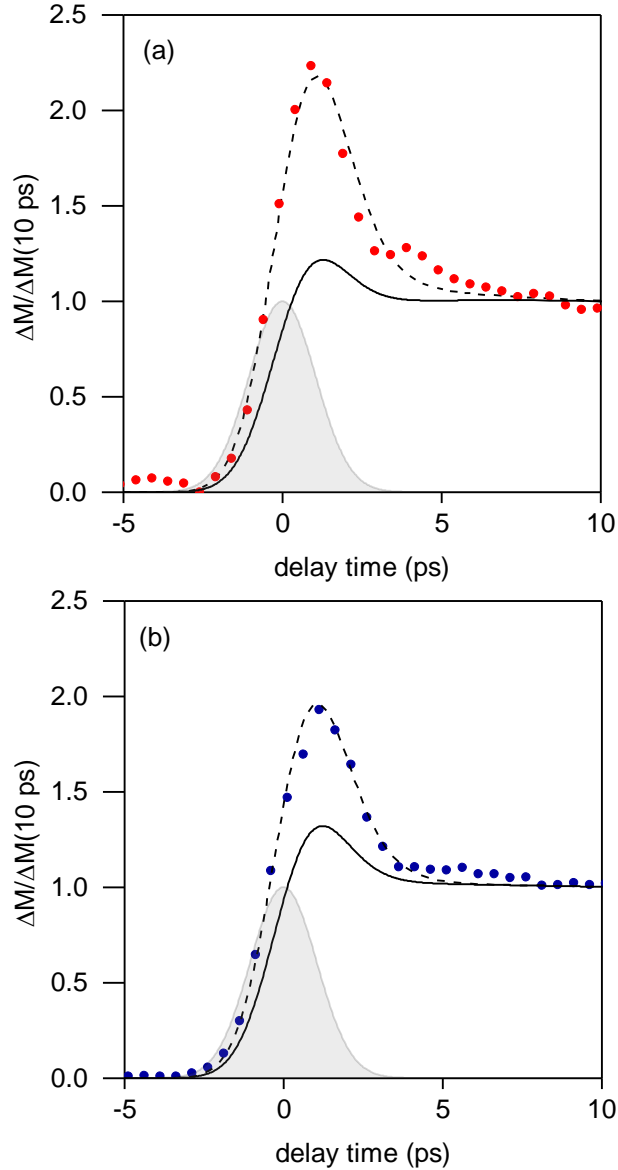

**Figure 2.** Optically induced demagnetization of (a) Pt/CoPt and (b) CoPt samples. The markers are normalized TR-MOKE measurements of the demagnetization of the Pt/CoPt and CoPt samples following optical absorption of ~0.2 and 0.7 J m$^{-2}$. The shaded region represents the power vs. time of the irradiating laser. The solid lines are a three-temperature model prediction for the magnetization dynamics with an electron phonon energy transfer coefficient predicted by scattering theory, $g_{ep} = 15 \times 10^{17}$ W m$^{-3}$ K$^{1}$. To explain the demagnetization data on picosecond scales with a thermal model, the net energy-transfer coefficient must be reduced from the theoretical value to $g_{ep} = 7 \times 10^{17}$ W m$^{-3}$ K$^{1}$, see dashed lines.

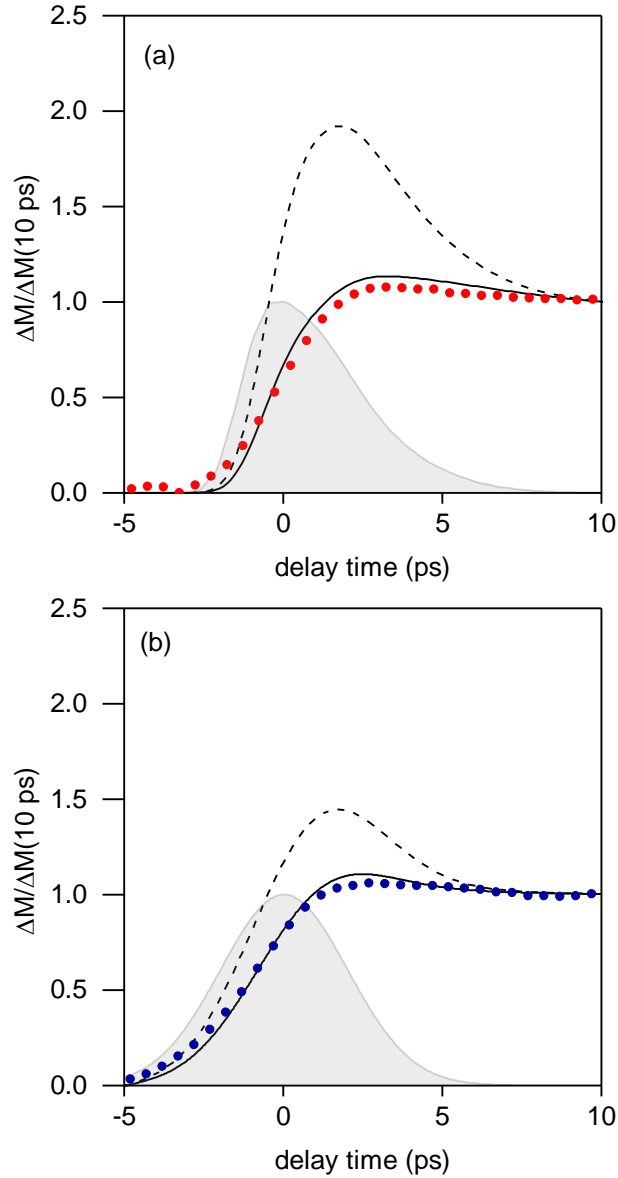

**Figure 3.** Electrically induced demagnetization in (a) Pt/CoPt and (b) CoPt samples. Markers are normalized TR-MOKE measurements of the magnetization of the samples after heating by an electrical pulse with ~0.5 A peak amplitude. The shaded region represents the power profile of the electrical pulse, as deduced from Protemics probe measurements. The heating time-scales are slightly different for the two samples due to differences in the Auson switch devices. The solid lines are three-temperature model predictions for the demagnetization with the theoretically calculated value of $g_{ep} = 15 \times 10^{17}$ W m$^{-3}$ K$^{-1}$. For comparison with the optical experiments, the dashed lines show predictions with a reduced electron-phonon energy transfer coefficient of $g_{ep} = 7 \times 10^{17}$ W m$^{-3}$ K$^{-1}$. A higher rate of energy transfer between electrons and phonons in the electrical heating experiments explains why there is no recovery of the magnetization in the picoseconds following electrical heating.

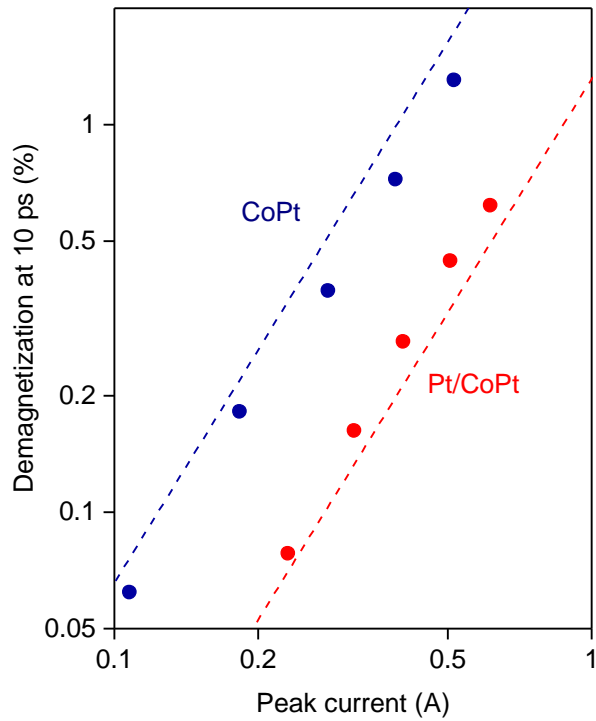

**Figure 4.** Demagnetization versus amplitude of the current pulse. Markers are TR-MOKE measurements of the demagnetization 10 ps after the electrical pulse heats the Pt/CoPt (red) and CoPt (blue). The dashed line are the predictions of our thermal model, using the value of $(1/M)dM/dT$ derived from optical demagnetization experiments and our calculation of the energy absorbed by the electrons via joule heating.

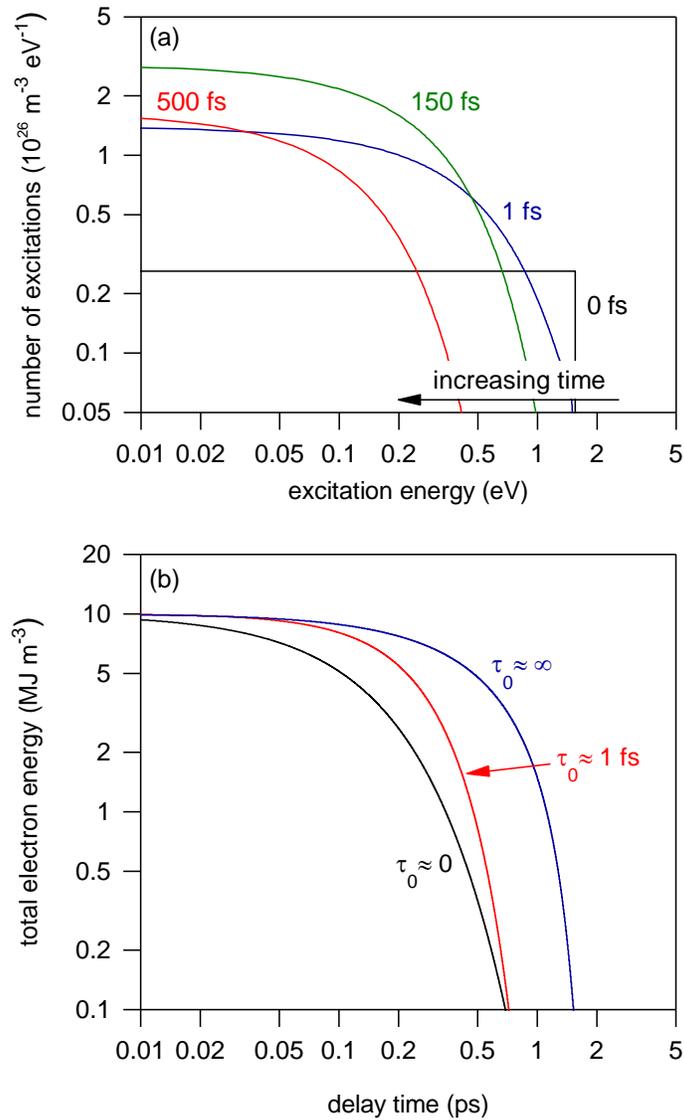

**Figure 5.** Nonthermal model predictions for the evolution of the electron energy in time. (a) Relaxation of the excited distributions due to electron-phonon and electron-electron scattering with $\tau_{ep} = 20 \text{ fs}$ and $\tau_0 = 1 \text{ fs}$. (b) Total energy stored by electronic excitations vs. time following impulsive heating for no electron-electron scattering ($\tau_0 = \infty$), strong electron-electron scattering ($\tau_0 = 0$), and realistic electron-electron scattering rates for Pt ($\tau_0 = 1 \text{ fs}$).

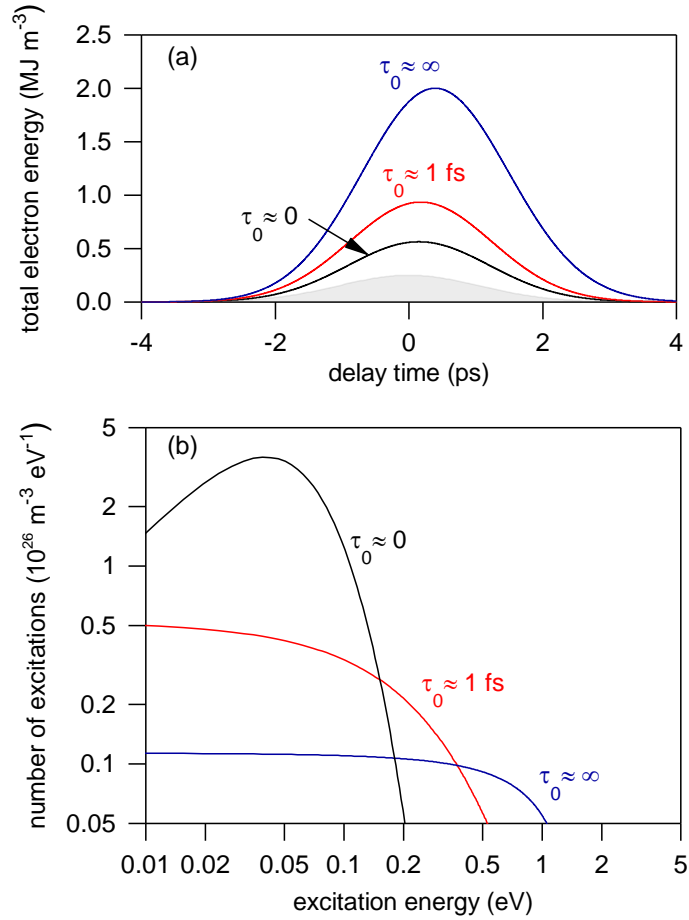

**Figure 6.** Nonthermal model predictions for the evolution of the electron energy in response to a 2.6 ps optical heat pulse. (a) Total energy stored by electronic excitations vs. time for no electron-electron scattering ($\tau_0 = \infty$), strong electron-electron scattering ($\tau_0 = 0$), and realistic electron-electron scattering rates for Pt ($\tau_0 = 1 \text{ fs}$). The gray region represents the temporal profile of the optical heating term, $A(t)$, with arbitrary units. (b) Predictions for the number of excitations across various excitation energies at zero time-delay. Despite the picosecond time-scale of the heating, the distribution of energy predicted by Eq. 3 is nonthermal for finite or zero electron-electron scattering.

**Acknowledgements**

This work was primarily supported by the Director, Office of Science, Office of Basic Energy Sciences, Materials Sciences and Engineering Division, of the U.S. Department of Energy under Contract No. DE-AC02-05-CH11231 within the Nonequilibrium Magnetic Materials Program (MSMAG). We also acknowledge the National Science Foundation Center for Energy Efficient Electronics Science for providing most of the experimental equipment and partially supporting the fabrication of samples.